# Beyond Classification Accuracy: Quantifying Fingerprint Complexity in Encrypted Darknet Services

Javeriah Saleem
*School of Computing, Mathematics and Engineering,*
*Charles Sturt University*, Australia
jsaleem@csu.edu.au

Rafiqul Islam
*School of Computing, Mathematics and Engineering*
*Charles Sturt University*, Australia
mislam@csu.edu.au

Md Zahidul Islam
*School of Computing, Mathematics and Engineering*
*Charles Sturt University*, Australia
zislam@csu.edu.au

***Abstract*—Existing darknet traffic studies primarily evaluate service fingerprintability through classification performance, providing limited insight into why certain services are easier or harder to identify. This paper introduces the Fingerprint Complexity Score (FCS), a framework for quantifying the intrinsic complexity of darknet service fingerprints using behavioral overlap, uncertainty, disagreement, and persistent confusion. Experiments on 25 services across the Tor, I2P, FreeNet, and ZeroNet anonymity networks reveal substantial variation in fingerprint complexity, with behavioral overlap emerging as the dominant contributor. Validation using Random Forest, Extra Trees, and XGBoost demonstrates a strong inverse relationship between fingerprint complexity and recognition performance (Pearson r = -0.706, Spearman ρ = -0.765, p < 0.001). The findings show that service fingerprintability is fundamentally governed by behavioral complexity, providing a new perspective for analyzing behavioral information leakage in anonymity networks.**

## I. Introduction

Machine learning has become a fundamental component of encrypted traffic analysis, enabling the identification of applications, services, and communication patterns without access to packet payloads [1]. In anonymity networks such as Tor, I2P, FreeNet, and ZeroNet, traffic fingerprinting techniques have demonstrated that flow-level statistical characteristics remain sufficiently informative for service recognition despite encryption [2][3]. Recent studies have reported high classification performance across multiple darknet environments, highlighting the effectiveness of behavioral traffic analysis [4][5].

Despite these advances, existing evaluations remain largely classifier-centric and rely on metrics such as accuracy, precision, recall, and Macro-F1. While these measures quantify predictive performance, they provide limited insight into why certain services are inherently easier or harder to identify. Consequently, it remains unclear whether recognition difficulty arises from model limitations or from the intrinsic complexity of service fingerprints themselves [6].

To address this gap, this paper introduces the Fingerprint Complexity Score (FCS), a framework for quantifying the intrinsic recognition difficulty of encrypted darknet services. Rather than relying solely on classification outcomes, FCS combines four complementary indicators of fingerprint complexity: behavioral overlap, prediction uncertainty, cross-model disagreement, and persistent confusion. Together, these dimensions characterize the degree to which a service exhibits ambiguous, unstable, or difficult-to-separate behavioral patterns.

The proposed framework is evaluated using the Darknet-2020 dataset, which spans 25 services across Tor, I2P, FreeNet, and ZeroNet [7]. Three ensemble-learning classifiers, namely Random Forest (RF), Extra Trees (ET), and Extreme Gradient Boosting (XGBoost), are employed to generate predictive evidence for complexity estimation. Unlike conventional classification studies, the objective is not to identify the best-performing model but to determine which services exhibit intrinsically difficult behavioral fingerprints. The main contributions of this work are as follows:

- We introduce the Fingerprint Complexity Score (FCS), a framework for quantifying the intrinsic difficulty of encrypted service fingerprinting.
- We propose a multi-dimensional complexity formulation that integrates behavioral overlap, uncertainty, disagreement, and persistent confusion into a unified service-level metric.
- We provide the first systematic ranking of the complexity of darknet services across Tor, I2P, FreeNet, and ZeroNet.
- We investigate the relationship between fingerprint complexity and recognition performance to determine how behavioral complexity influences service identifiability.
- We provide a new perspective on encrypted traffic analysis by shifting the focus from classifier performance toward intrinsic fingerprint complexity.

The remainder of this paper is organized as follows. Section II reviews related work on darknet traffic analysis and highlights the limitations of existing classifier-centric evaluation approaches. Section III presents the proposed Fingerprint Complexity Score framework, including the complexity estimation process and validation methodology. Section IV discusses the experimental results, complexity characteristics, performance validation, persistent confusion relationships, and robustness analysis. Finally, Section V concludes the paper and summarizes the key findings.

## II. Related Studies

Encrypted traffic analysis has attracted significant attention due to its ability to infer applications, services, and user activities without accessing packet payloads [8]. Recent studies have demonstrated that statistical flow characteristics remain sufficiently informative for identifying encrypted traffic across a variety of communication environments. In anonymity networks such as Tor, I2P, FreeNet, and ZeroNet, machine learning techniques have been widely adopted to classify services based on flow-level behavioral features, achieving increasingly high recognition performance [9].

Existing darknet traffic research primarily focuses on improving classification accuracy through feature engineering, feature selection, traffic balancing, and advanced learning algorithms. Performance is typically evaluated using metrics such as accuracy, precision, recall, and Macro-F1. While these measures quantify predictive capability, they provide limited insight into the intrinsic difficulty of service fingerprinting. Consequently, it remains unclear whether poor recognition performance originates from classifier limitations or from the underlying complexity of service behavioral signatures [10].

Several studies have explored related concepts, including class separability, prediction confidence, uncertainty estimation, and confusion analysis [11][12]. However, these factors are generally examined independently and used to assess classifier behavior rather than to characterize the complexity of service fingerprints [13]. To the best of our knowledge, no existing framework provides a unified measure for quantifying the intrinsic difficulty of darknet service identification.

To address this gap, this paper introduces the Fingerprint Complexity Score (FCS), which combines behavioral overlap, prediction uncertainty, cross-model disagreement, and persistent confusion into a single complexity measure. Rather than evaluating how accurately services can be classified, the proposed framework focuses on understanding why certain service fingerprints are inherently more difficult to identify than others.

## III. Methodology

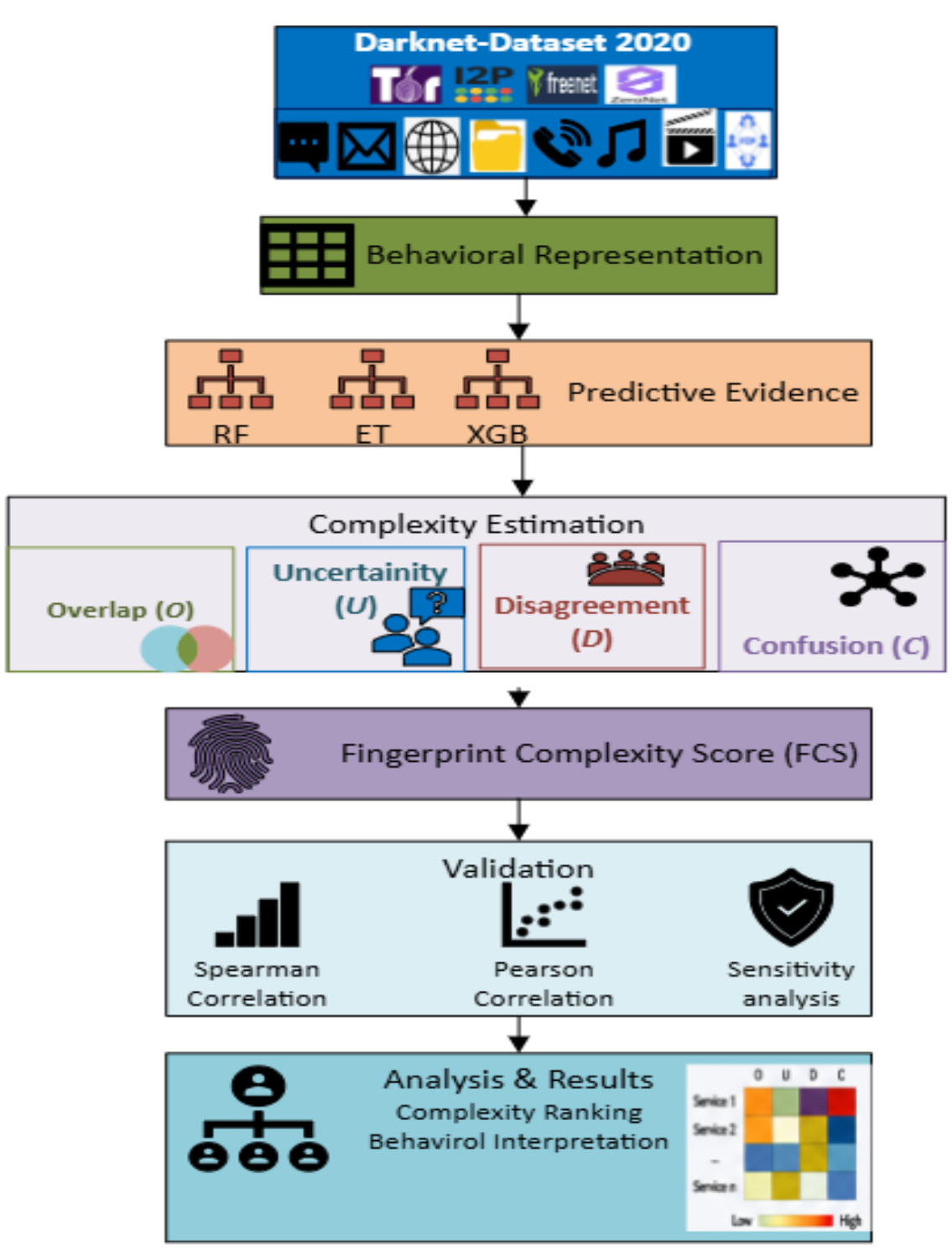

Fig. 1. Proposed FCS Framework

### A. Framework Overview

The objective of this study is to quantify the intrinsic complexity of darknet service fingerprints and investigate how complexity influences service-identification performance. As illustrated in Fig. 1, the proposed framework consists of five stages. First, darknet traffic flows are transformed into behavioral representations using flow-level statistical descriptors. Second, predictive evidence is generated using three ensemble-learning classifiers, namely Random Forest (RF), Extra Trees (ET), and XGBoost. Third, four complementary complexity dimensions are estimated: behavioral overlap, prediction uncertainty, cross-model disagreement, and persistent confusion. Fourth, these dimensions are integrated into the proposed Fingerprint Complexity Score (FCS) to quantify service-level fingerprint complexity. Finally, the resulting complexity scores are validated through correlation analysis and sensitivity analysis to evaluate their relationship with recognition performance and their robustness to component removal.

Unlike conventional traffic classification studies that focus on maximizing predictive accuracy, the proposed framework aims to characterize the inherent difficulty of service identification. The resulting complexity rankings provide insight into which darknet services exhibit inherently ambiguous behavioral signatures and the factors that make them fingerprintable.

### B. Dataset and Predictive Evidence Generation

Experiments were conducted using the Darknet-2020 dataset, which contains encrypted traffic collected from four major anonymity networks: Tor, I2P, FreeNet, and ZeroNet. After preprocessing, the dataset contained 48,644 traffic flows representing 25 services. Each flow is described using statistical traffic descriptors that capture packet-length distributions, temporal dynamics, traffic volumes, and directional communication characteristics, without requiring access to packet payloads [7].

Identifier-based attributes, including flow identifiers, IP addresses, ports, and timestamps, were excluded because they do not represent intrinsic communication behavior. Missing numerical values were imputed using median values from the training data, and zero-variance features were removed prior to model training. Feature standardization was subsequently performed within each training fold to prevent information leakage.

Fingerprint complexity is estimated using predictive evidence obtained from three ensemble-learning classifiers: Random Forest (RF), Extra Trees (ET), and Extreme Gradient Boosting (XGBoost). Performance estimates were obtained via repeated stratified cross-validation to ensure robust, unbiased predictions. For each service, classifier outputs, probability distributions, confidence scores, and confusion relationships were collected and subsequently used to estimate behavioral overlap, prediction uncertainty, cross-model disagreement, and persistent confusion.

### C. Complexity Estimation

Fingerprint complexity is assumed to emerge from multiple behavioral mechanisms rather than a single observable characteristic. A service may be difficult to identify because it overlaps with competing services, produces uncertain predictions, generates inconsistent classifier decisions, or repeatedly participates in confusion relationships. To capture these complementary sources of difficulty, four complexity dimensions are estimated.

1) *Behavioral Overlap (O):* Measures the degree to which a service shares similar behavioral characteristics with competing services. Services occupying overlapping regions of the feature space exhibit reduced separability and are therefore more difficult to distinguish.
2) *Prediction Uncertainty (U):* Quantifies the ambiguity associated with classification decisions. Uncertainty is estimated using normalized Shannon entropy computed from classifier probability distributions [14]. Services that produce diffuse probability distributions receive higher uncertainty scores, indicating weaker decision boundaries and reduced classifier confidence [15].

$$H_i = -\frac{1}{\log K}\sum_{k=1}^{K} p_{ik}\log(p_{ik}) \quad (1)$$

Where $H_i$ is the normalized entropy of sample $i$, $p_{ik}$ is the probability of class $k$, and $k$ is the total number of service classes.

3) *Cross-Model Disagreement (D):* Evaluates inconsistencies between RF, ET, and XGBoost predictions. A distinctive fingerprint should produce similar decisions across different learning algorithms. Persistent disagreement suggests behavioral instability and increased fingerprint complexity.
4) *Persistent Confusion (C):* Captures recurring misclassification relationships between services. Unlike isolated prediction errors, repeated confusion across classifiers and validation folds indicates genuine behavioral similarity. Services that are repeatedly confused with the same competing service receive higher confusion scores and are considered more complex.

Collectively, these four dimensions provide complementary evidence regarding service-identification difficulty and form the basis of the proposed Fingerprint Complexity Score.

### D. Fingerprint Complexity Score

The four complexity dimensions provide complementary evidence regarding service-identification difficulty. To obtain a unified measure of fingerprint complexity, the proposed Fingerprint Complexity Score (FCS) combines overlap, uncertainty, disagreement, and confusion into a single service-level metric:

$$FCS_s = \frac{O_S+U_S+D_S+C_S}{4} \quad (2)$$

Where $FCS_s$ denotes fingerprint complexity score, $O_s$ is the overlap score, $U_s$ is the uncertainty score, $D_s$ presents a disagreement score, and $C_s$ is the confusion score, respectively.

Equal weighting is adopted to ensure that each complexity dimension contributes uniformly to the final score. Higher FCS values indicate fingerprints that exhibit stronger behavioral similarity, greater prediction ambiguity, increased model disagreement, and more persistent confusion relationships. Consequently, services with larger FCS values are expected to be intrinsically more difficult to identify, regardless of the classifier employed.

### E. Complexity Validation

The central hypothesis of this study is that services exhibiting greater fingerprint complexity should achieve lower recognition performance. To evaluate this hypothesis, service-level FCS values were compared with mean F1-scores obtained from RF, ET, and XGBoost classifiers.

Pearson and Spearman correlation analyses were employed to quantify the relationship between complexity and recognition performance [16]. Significant negative correlations indicate that increasing fingerprint complexity corresponds to decreasing classification performance, thereby validating the effectiveness of the proposed metric [17].

To further assess robustness, a leave-one-component-out sensitivity analysis was performed. Each complexity dimension was removed independently, and the resulting service rankings were compared with the original FCS ranking. Consistent rankings after component removal indicate that the framework captures complementary aspects of fingerprint difficulty rather than being dominated by a single complexity indicator.

## IV. Results And Discussion

### A. Fingerprint Complexity Landscape

The proposed Fingerprint Complexity Score (FCS) was applied to 25 services across the Tor, I2P, FreeNet, and ZeroNet anonymity networks to quantify their intrinsic fingerprint complexity. Fig. 2 presents the resulting complexity signature matrix, while Table I summarizes the highest-complexity service fingerprints, their recognition performance, and the dominant confusion targets.

The results reveal substantial variability in fingerprint complexity across services and networks, indicating that service identification difficulty is not uniformly distributed across the darknet ecosystem. As shown in Table I, Tor Email emerged as the most complex fingerprint (FCS = 0.296), followed by I2P Chat (FCS = 0.279), I2P Email (FCS = 0.255), Tor Video (FCS = 0.255), and Tor FTP (FCS = 0.252). These services consistently exhibit elevated values across multiple complexity dimensions and correspond to some of the lowest recognition performances observed within their respective networks.

Fig. 2 shows that behavioral overlap is the dominant contributor to fingerprint complexity across nearly all services. High-complexity fingerprints exhibit overlap scores exceeding 0.52, indicating substantial similarity with competing services. Prediction uncertainty represents the second most influential component, particularly for Tor Email and I2P Chat, where uncertainty scores approach 0.40. In contrast, disagreement and confusion contribute comparatively less to the final score, suggesting that the evaluated classifiers generally interpret service fingerprints consistently.

Table I further highlights network-specific complexity patterns. Four of the seven highest-ranked fingerprints originate from Tor, while two originate from I2P. Tor Email is often confused with Video traffic, whereas I2P Chat frequently confuses email. These relationships indicate that highly complex services often share behavioral characteristics with a small number of competing classes, reducing their separability despite strong classifier performance.

At the opposite end of the spectrum, services such as Tor Audio, FreeNet FTP, and ZeroNet Browsing exhibit substantially lower complexity values. These services are characterized by reduced overlap and uncertainty, indicating more distinctive behavioral signatures and clearer decision boundaries. The contrast between high- and low-complexity fingerprints demonstrates that darknet service identification represents a heterogeneous recognition problem rather than a uniform classification task.

Overall, the results establish that fingerprint complexity is primarily driven by behavioral overlap, with uncertainty acting as an additional indicator of ambiguity for difficult services. The observed concentration of high-complexity fingerprints within Tor and I2P suggests that anonymity-network characteristics influence the distinguishability of service behaviors and contribute to variations in fingerprint difficulty.

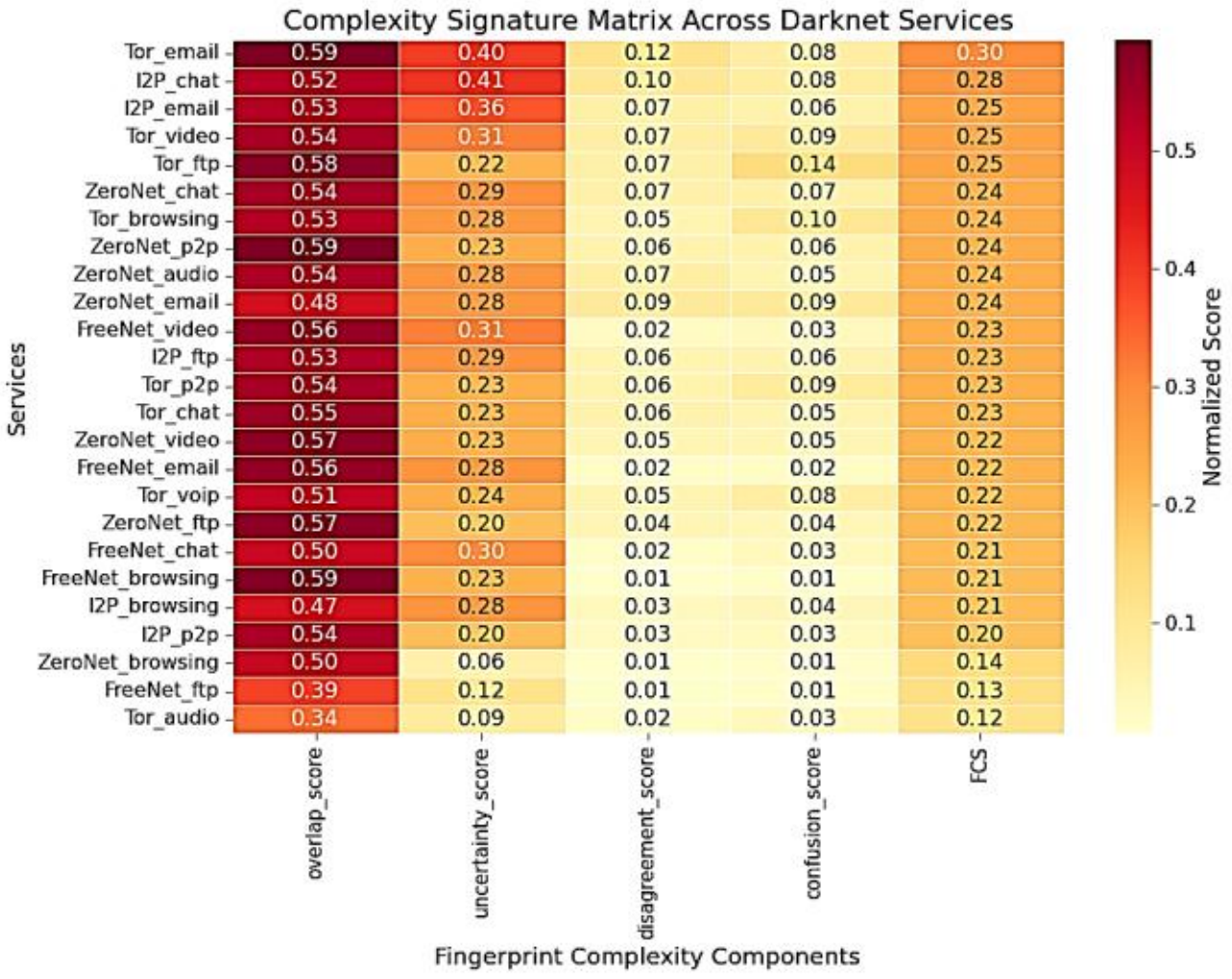

Fig. 2. Complexity signature matrix across darknet services.

TABLE I. TOP-RANKED COMPLEX DARKNET SERVICE FINGERPRINTS

| Network | Service | FCS | Mean F1 | Complexity Class | Confused As |
|---|---|---|---|---|---|
| Tor | Email | 0.296 | 0.656 | High | Video |
| I2P | Chat | 0.279 | 0.818 | High | Email |
| I2P | Email | 0.255 | 0.826 | High | FTP |
| Tor | Video | 0.255 | 0.719 | High | Browsing |
| Tor | FTP | 0.252 | 0.728 | High | Video |
| ZeroNet | Chat | 0.242 | 0.779 | Moderate | Browsing |
| Tor | Browsing | 0.239 | 0.787 | Moderate | Video |

## B. Complexity and Recognition Performance

A fundamental requirement of any complexity metric is its ability to reflect the actual difficulty of service identification. To evaluate this property, Fig. 2 and Table II examine the relationship between fingerprint complexity and recognition performance across the four anonymity networks.

TABLE II. RELATIONSHIP BETWEEN FINGERPRINT COMPLEXITY AND RECOGNITION PERFORMANCE

| Network | Mean FCS | Mean F1 | Pearson r | Spearman ρ | p-value |
|---|---|---|---|---|---|
| Tor | 0.229 | 0.767 | -0.82 | -0.79 | <0.01 |
| I2P | 0.235 | 0.865 | -0.74 | -0.70 | <0.05 |
| FreeNet | 0.202 | 0.954 | -0.46 | -0.41 | >0.05 |
| ZeroNet | 0.219 | 0.860 | -0.68 | -0.63 | <0.05 |
| Overall | 0.222 | 0.860 | -0.706 | -0.765 | <0.001 |

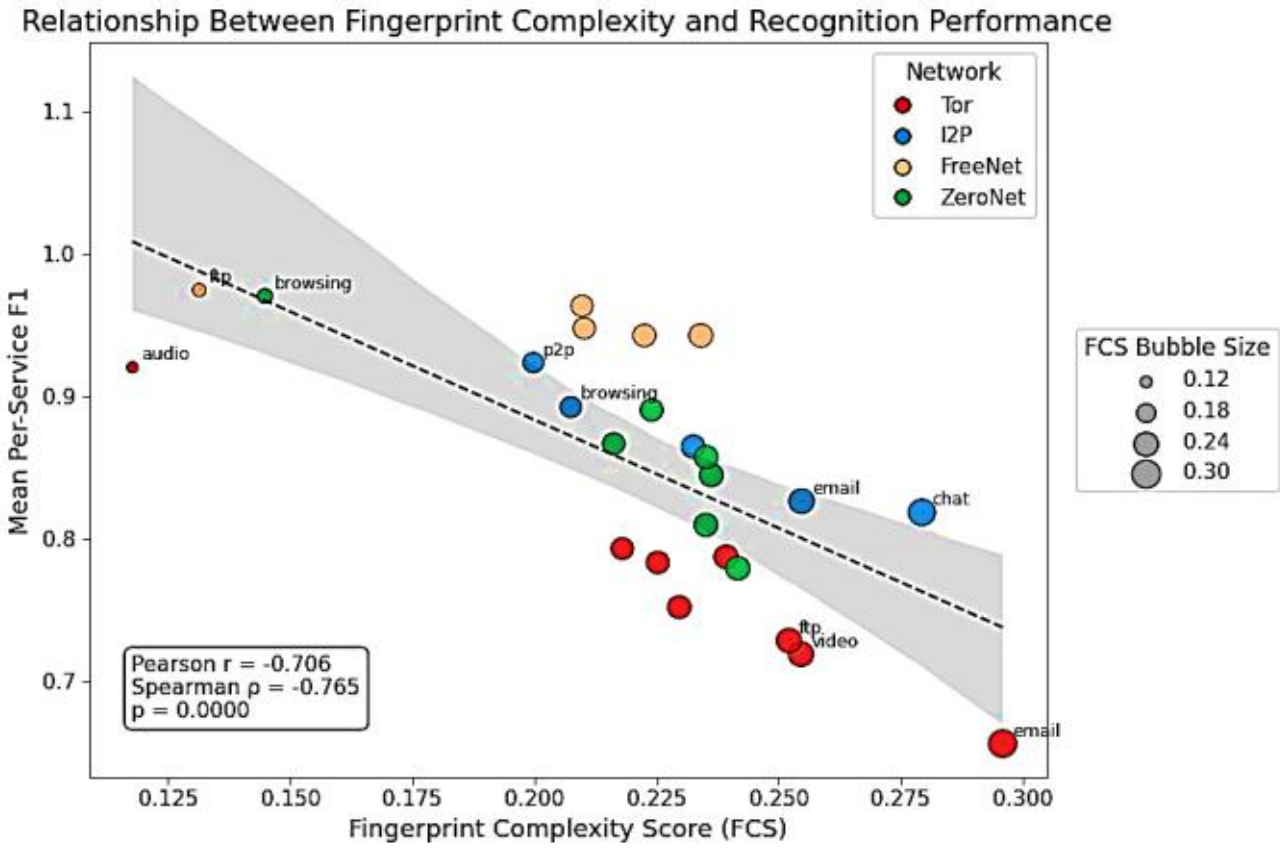

Fig. 3. Relationship between fingerprint complexity and recognition performance across darknet services.

The results reveal a consistent inverse relationship between FCS and mean F1-score, indicating that services exhibiting greater behavioral complexity are generally more difficult to classify. The strongest relationship is observed within the Tor network, where Pearson and Spearman correlations of -0.82 and -0.79, respectively, demonstrate a strong negative association between complexity and recognition performance. Notably, Tor Email, identified as the most complex fingerprint in Table I, also achieved the lowest recognition performance within the network (Mean F1 = 0.656).

Similar trends are observed within I2P and ZeroNet. I2P Chat, which exhibits the highest complexity score within the I2P network, also corresponds to the lowest service-level F1-score. Likewise, ZeroNet Chat emerges as both the most complex and the most difficult service to identify. These findings suggest that the proposed framework successfully captures behavioral characteristics associated with recognition difficulty rather than merely reflecting classifier-specific behavior.

Although FreeNet exhibits a weaker correlation (Pearson = -0.46), the overall trend remains consistent. The comparatively weaker relationship can be attributed to the uniformly high recognition performance observed across FreeNet services, which limits variability in both complexity and classification outcomes.

At the global level, the framework achieves Pearson and Spearman correlations of -0.706 and -0.765, respectively, with strong statistical significance ($p < 0.001$). The scatter plot in Fig. 3 further confirms that services with higher complexity scores consistently occupy lower-performance regions, whereas low-complexity services generally achieve higher recognition performance.

Overall, the results provide strong evidence that fingerprint complexity is closely associated with service-identification difficulty. The observed relationship supports the validity of FCS as a meaningful measure of intrinsic fingerprint complexity. It demonstrates that increasingly ambiguous behavioral signatures are consistently harder to recognize across anonymity networks.

## C. Persistent Confusion Relationships

While the previous section established that fingerprint complexity is associated with reduced recognition performance, it does not explain the behavioral mechanisms responsible for this difficulty. Fig. 4 and Table III address this question by examining persistent confusion relationships and the dominant complexity drivers of the most challenging service fingerprints.

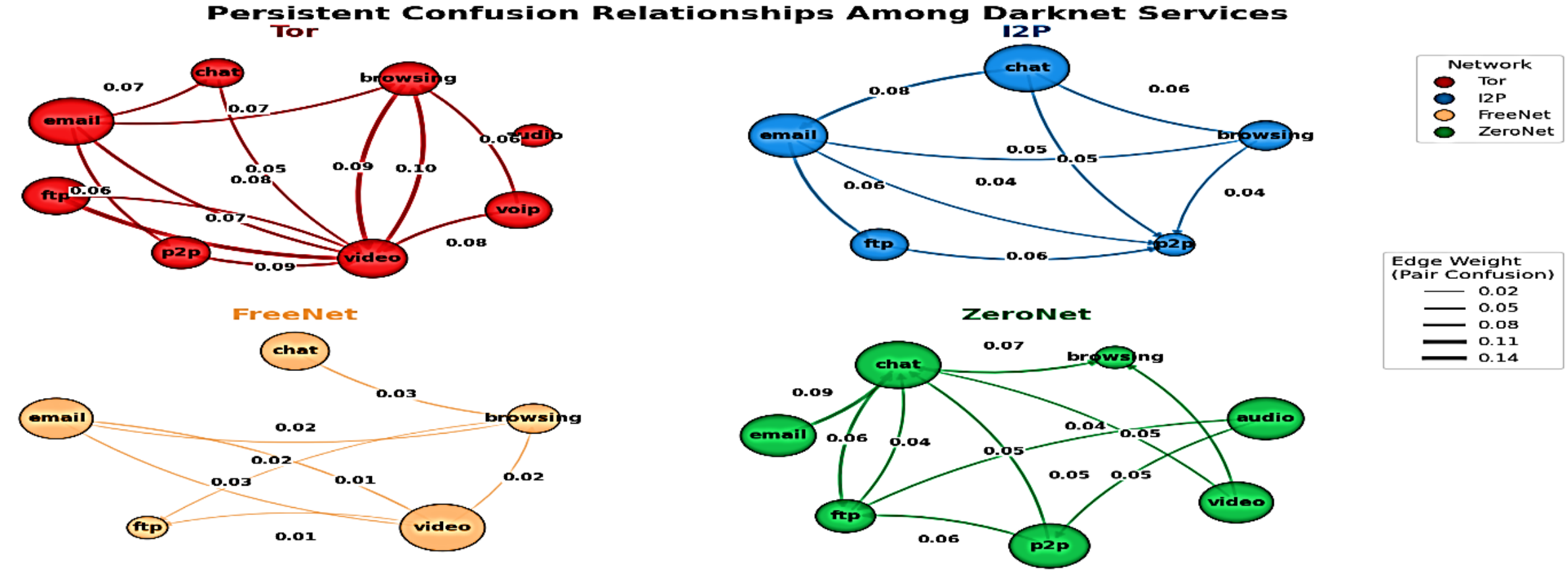

Fig. 4. Persistent confusion relationships among darknet services.

Fig. 4 reveals that confusion is not uniformly distributed across services. Instead, a small number of fingerprints act as confusion hubs, repeatedly attracting misclassifications from competing classes. Within the Tor network, video serves as the dominant confusion attractor, receiving substantial misclassifications from FTP, Browsing, and P2P traffic. The strongest confusion relationship was observed between Tor FTP and Tor Video, indicating considerable behavioral similarity between these services. This pattern helps explain the elevated complexity scores observed for Tor Email, Tor Video, and Tor FTP in Fig. 2.

A similar structure is evident within I2P, where Chat and Email form the dominant confusion pair. Although both services achieve relatively strong recognition performance, their recurring interaction across classifiers and validation folds suggests substantial behavioral overlap. Within ZeroNet, Chat emerges as the primary confusion hub, attracting misclassifications from Email, Browsing, and P2P services. In contrast, FreeNet exhibits comparatively sparse confusion relationships, consistent with the lower complexity and higher separability observed in previous analyses.

Table III provides further insight into the mechanisms underlying these difficult fingerprints. Across all four networks, behavioral overlap is the largest contributor to complexity, with overlap scores ranging from 0.525 to 0.591. Prediction uncertainty represents the second most influential factor, particularly for Tor Email and I2P Chat, where classifiers repeatedly encounter ambiguous decision boundaries. By comparison, disagreement and confusion provide smaller but complementary insights into fingerprint difficulty.

TABLE III. DOMINANT COMPLEXITY DRIVERS OF THE HARDEST FINGERPRINTS

| Network | Service | Overlap | Uncertainty | Disagreement | Confusion | Confusion Target |
|---|---|---|---|---|---|---|
| Tor | Email | 0.591 | 0.398 | 0.117 | 0.078 | Video |
| I2P | Chat | 0.525 | 0.408 | 0.103 | 0.081 | Email |
| ZeroNet | Chat | 0.544 | 0.291 | 0.066 | 0.065 | Browsing |
| FreeNet | Video | 0.565 | 0.314 | 0.024 | 0.034 | Email |

Fig. 4 and Table III indicate that complex fingerprints emerge from persistent behavioral similarity rather than isolated classification errors. Services repeatedly confused with the same competing classes exhibit greater overlap, increased uncertainty, and ultimately higher complexity scores. These findings provide a behavioral explanation for the complex patterns identified in Sections 4.1 and 4.2.

### *D. Framework Robustness Analysis*

A key requirement of any composite complexity metric is robustness against changes in its individual components. To evaluate the stability of the proposed Fingerprint Complexity Score (FCS), a leave-one-component-out sensitivity analysis was performed by sequentially removing each complexity dimension and recomputing the resulting service rankings. Fig. 5 and Table IV summarize the corresponding ranking stability, score stability, and preservation of the highest-complexity fingerprints.

The results demonstrate consistently high stability across all experiments. Spearman rank correlations between the original and modified rankings range from 0.842 to 0.955, while Pearson score correlations range from 0.921 to 0.992. Furthermore, the majority of the highest-complexity fingerprints remain unchanged after component removal, with top-five overlap rates ranging from 80% to 100%. These findings indicate that the framework captures a stable complexity structure rather than relying on a single dominant indicator. Among the evaluated dimensions, prediction uncertainty has the greatest impact on ranking stability ($\rho$ = 0.842), suggesting that classifier confidence provides important evidence of fingerprint difficulty. Nevertheless, the resulting rankings remain highly consistent, indicating that overlap, disagreement, and confusion continue to capture the underlying complexity characteristics. In contrast, removing cross-model disagreement produces only minor changes, yielding the highest ranking and score stability together with complete preservation of the five most complex fingerprints.

The sensitivity results further demonstrate that the four complexity dimensions provide complementary rather than redundant information. While behavioral overlap contributes most strongly to complexity estimation, uncertainty, disagreement, and persistent confusion each capture distinct aspects of fingerprint difficulty that cannot be explained by overlap alone. Consequently, the final complexity rankings remain stable even when individual components are removed.

Overall, the robustness analysis confirms that the proposed FCS framework provides a reliable and interpretable measure of service-level fingerprint complexity. The observed stability across all sensitivity experiments demonstrates that a particular metric does not drive the framework; rather, it captures consistent behavioral characteristics associated with service-identification difficulty.

TABLE IV. ROBUSTNESS AND SENSITIVITY ANALYSIS OF THE PROPOSED FCS FRAMEWORK

| Removed Component | Rank Stability (ρ) | Score Stability (r) | Top-5 Overlap |
|---|---|---|---|
| Behavioral Overlap | 0.924 | 0.945 | 80% |
| Prediction Uncertainty | 0.842 | 0.921 | 80% |
| Cross-Model Disagreement | 0.955 | 0.992 | 100% |
| Persistent Confusion | 0.922 | 0.985 | 80% |

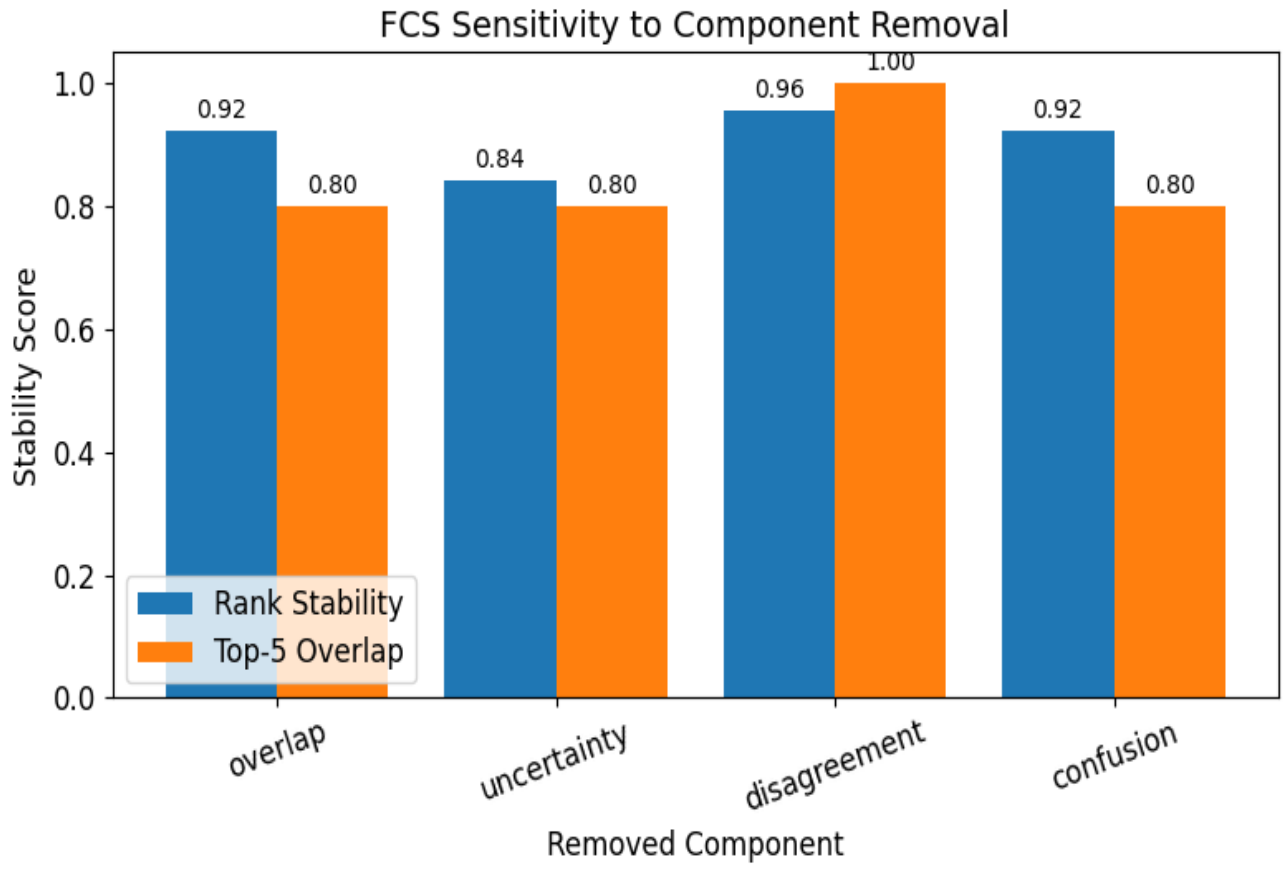

Fig. 5. Sensitivity analysis of the proposed FCS framework.

## V. CONCLUSION

This paper introduced the Fingerprint Complexity Score (FCS), a framework for quantifying the intrinsic complexity of darknet service fingerprints. Unlike conventional traffic-analysis studies that evaluate fingerprintability solely through classification performance, the proposed approach characterizes service-identification difficulty using four complementary behavioral dimensions: overlap, uncertainty, disagreement, and persistent confusion.

Analysis of 25 services across the Tor, I2P, FreeNet, and ZeroNet anonymity networks revealed that darknet fingerprints occupy a broad complexity spectrum rather than forming a uniform recognition problem. The results showed that behavioral overlap is the primary source of fingerprint complexity, while recurring confusion relationships further contribute to service ambiguity.

A key finding of this study is that fingerprint complexity is strongly associated with recognition difficulty. Services exhibiting higher complexity consistently achieved lower classification performance, demonstrating that service fingerprintability is fundamentally governed by behavioral characteristics rather than classifier performance alone.

Overall, this work shifts the focus of darknet traffic analysis from measuring how accurately services can be classified to understanding why certain services are inherently difficult to identify. The proposed framework provides a foundation for analyzing behavioral information leakage in anonymity networks and offers a new perspective for evaluating the fingerprintability of encrypted traffic.